\documentclass[12pt,preprint]{aastex}
\usepackage{natbib}
\usepackage{amsfonts}
\usepackage{color}
\usepackage{ulem}
\usepackage{amsmath}
\usepackage{mathtools}
\usepackage{pdflscape}
\usepackage{graphicx}
\usepackage[export]{adjustbox}
\usepackage{multirow}

\newcommand{\rsun}{$\,R_{S}$}
\newcommand{\degrees}{$^{\circ}$}
\newcommand{\thetabn}{$\theta_{Bn}$~}

\shorttitle{Early-Stage SEP Acceleration}
\shortauthors{K. A. Kozarev, M. A. Dayeh, A. Farahat}

\begin{document}

\title{Early-Stage SEP Acceleration by CME-Driven Shocks with Realistic Seed Spectra: I. Low corona\\}

\author{Kamen A. Kozarev\altaffilmark{1}, Maher A. Dayeh\altaffilmark{2}, Ashraf Farahat\altaffilmark{3}}
\altaffiltext{1}{Institute of Astronomy and National Astronomical Observatory, Bulgarian Academy of Sciences, Bulgaria}
\altaffiltext{2}{Southwest Research Institute, TX, USA}
\altaffiltext{3}{King Fahd University of Petroleum and Minerals, Saudi Arabia}

\begin{abstract}
An outstanding problem in heliospheric physics is understanding the acceleration of solar energetic particles (SEP) in coronal mass ejections (CMEs) and flares. A fundamental question is whether the acceleration occurs in interplanetary space, or near the Sun. Recent work has shown that CME-driven shocks may produce SEPs while still below 5 solar radii. In this work we explore SEP acceleration during the onset of CMEs and shocks even lower in the corona, using realistic suprathermal spectra, for a selection of events. We have calculated quiet-time, pre-event suprathermal particle spectra from 1 AU observations, and scaled them back to the low corona to serve as seed spectra. For each event, AIA observations and the CASHeW framework were used to model the compressive/shock wave kinematics and its interaction with the corona. The proton acceleration was then modeled using an analytic diffusive shock acceleration model as the shock waves propagate between $\sim$1.05 and $\sim$1.3 solar radii. We demonstrate the capability of low coronal shock-related EUV waves to accelerate protons to multi-MeV energies in a matter of minutes, in the very early stages of the associated solar eruptions. We find that strong proton energization occurs for high values of the density jump, Alfv\'en Mach number, and shock speed. In future work the results of this early-stage shock acceleration will be used to model the continued acceleration higher in the corona.\end{abstract}

\section{Introduction}
\label{s1}

Coronal Mass Ejections (CMEs), are the most energetic manifestations of short-term solar activity, and a complex phenomenon with multiple manifestations. They occur throughout the solar cycle, and consist of hot coronal matter and magnetic field, ejected in impulsive periods of magnetic energy release, centered on complex active regions (AR) on the Sun's surface. CMEs are often accompanied by flares, which constitute the release of high-energy electromagnetic radiation. The large amount of kinetic energy of the ejecta is such, that they often drive shock waves immediately after their onset, and as low as 1.2\rsun, as evidenced by observations of coronal radio bursts \citep{Gopalswamy:2011, Carley:2013}. 

In extreme ultraviolet (EUV) light observations, a related phenomenon - the so-called EUV waves or EIT waves \citep{Thompson:1998} - has been observed over the past 20~years, which has been connected to the onset and early stages of CMEs. These large-scale, structured transient brightenings are best observed in wavelengths corresponding to solar coronal temperatures on the order of logT$_{cor}$=[5.8-6.3]. They have also been dubbed large-scale coronal propagating fronts \citep[LCPF]{Nitta:2013} or coronal bright fronts \citep{Long:2011}, to avoid calling them waves, due to the long-standing debate as to their physical nature. Recent focused investigations involving both observations and numerical modeling have confirmed that they are most likely magnetohydrodynamic waves or shocks \citep{Patsourakos:2012, Long:2017}. Indeed, due to the inhomogeneity of the corona and the steep radial gradient of Alfv\'en speed \citep{Evans:2008}, CBFs may switch dynamically between waves and shocks, depending on the local plasma environment. 

Detailed observations with high spatial and temporal resolution with the Atmospheric Imaging Assembly \citep[AIA]{Lemen:2012} imaging instrument on the Solar Dynamics Observatory spacecraft \citep[SDO]{Pesnell:2012} show that CBFs are driven by the impulsively expanding eruptive filaments or coronal magnetic loops of the CME \citep{Downs:2012}, including what is known as the CME bubble \citep{Patsourakos:2010}. Furthermore, previous studies have found a strong correspondence between CBFs and coronal shocks observed in radio bursts \citep{Kozarev:2011, Cunha-Silva:2015}. These results have raised the question of whether CME-driven MHD shocks can produce charged particle populations at solar energetic particle (SEP) energies ($>$10~MeV) very shortly after the CME onset in the corona. SEPs are an important topic of study in heliospheric physics, not only because they probe interplanetary magnetic fields, but also because they can present considerable radiation risk for space missions, both manned and unmanned. Several numerical modeling studies of realistic CME and particle acceleration simulations have suggested that indeed, shocks driven by fast CMEs in the low and middle corona (below 10\rsun) can produce SEPs with energies up to 1~GeV \citep{Sandroos:2009, Kozarev:2013, Schwadron:2015,Afanasiev:2018}.

Several authors recently studied the characteristics of CBFs and their capability to accelerate high-energy charged particles \citep{Kozarev:2011, Long:2011, Kozarev:2015, Rouillard:2016}. More recently, \citet{Long:2014} developed a framework for the characterization of the on-disk dynamics of multiple CBFs, while \citet{Kozarev:2017} put together a suite of tools for off-limb CBF characterization, currently within the field of view of AIA. Studying off-limb events allows to probe the structure of the front in coronal height, and thus to estimate its three-dimensional structure and possible interaction with coronal magnetic fields.

The ability of shocks to accelerate particles depends strongly on their relative speeds to the shock - only particles with sufficiently high momentum can enter the acceleration process \citep{Giacalone:2006}. This `injection momentum' depends on the particular shock properties in the region of interaction, such as the local upstream magnetic field strength, its small-scale fluctuations, and its relative orientation to the shock normal direction - the so-called angle $\theta_{BN}$. That is why suprathermal particles, with energies at least an order of magnitude beyond the typical thermal energies of the corona and solar wind (E$_sup > 10~keV$), are of prime interest as the source population for shock acceleration in the corona, as well as interplanetary space. Unfortunately, in situ observations of suprathermal solar charged particles can only be made near 1 AU currently, and so obtaining pre-event, quiet-time suprathermal observations from in situ data remains the only way to estimate the source populations of coronal shock-accelerated SEPs \citep{Dayeh:2017} somewhat reliably. This has previously been exploited by \citet{Kozarev:2013}, who used 1 AU, quiet-time, suprathermal Helium spectra, scaled back to the Sun, to drive a focused transport kinetic model coupled to an MHD model of a CME in the low and middle corona.

In this work, we combine the functionality of the CASHeW framework for characterizing low-coronal off-limb waves and shocks \citep{Kozarev:2017} with an analytic diffusive shock acceleration model \citep{Kozarev:2016}, and realistic suprathermal source spectra, in order to model the early-stage shock-driven SEP acceleration in 9 eruptions between 2011 and 2013. The spatial domain we focus on is within the field of view of AIA (up to $\sim$1.3\rsun), as we use the AIA observations to derive shock properties directly. In future work, we plan to extend the modeling domain and use the results presented here to investigate the acceleration higher in the corona, up to $\sim$6\rsun. The structure of the paper is as follows: Section \ref{s2} details the event selection and the observations used. Section \ref{s3} describes the CBF analysis. Section \ref{s4} describes the modeling of the shock parameters and particle acceleration. In Section \ref{s5} we discuss the results, while Section \ref{s6} provides a summary.

\section{Event Selection and Observations}
 \label{s2}

\subsection{Event Selection}
We selected 9 western events, which 1) included a quasi-spherical EUV coronal bright front (as observed in the SDO/AIA 193~/AA and 211~/AA channels), 2) had statistically-significant suprathermal quiet-time Oxygen spectra, 3) were not preceded by a solar event for at least a day, so that the quiet-time suprathermal populations dominate, 4) were associated with considerable increase in the high-energy Oxygen (O) fluxes during the early stages of the events. We present the main properties of the events in Table \ref{eventstable}.

\begin{table}[ht]
\centering
\begin{tabular}{c c c c c c c c}
\hline
Event \# & Date & CBF Start & Source Location & Flare Class & $Q_0$ & $Q_1$\\
\hline
E1 & 05/15/2011 & 23:30 & W44N09 & C4.8 & 3.27 & 1.58$\pm$0.22\\
E2 & 06/07/2011 & 06:20 & W44S21 & M2.5 & 2.71 & 1.61$\pm$0.06\\
E3 & 08/04/2011 & 03:50 & W38N20 & M9.3 & 3.29 & 1.46$\pm$0.27\\
E4 & 10/20/2011 & 03:05 & W88N20 & M1.6 & 3.74 & 1.60$\pm$0.69\\
E5 & 05/26/2012 & 20:30 & W89N24 & -- & 4.19 & 2.55$\pm$0.52\\
E6 & 11/19/2013 & 10:15 & W71S19 & X1.0 & 3.06 & 1.31$\pm$0.16\\
E7 & 12/07/2013 & 07:15 & W47S15 & M1.2 & 3.90 & 2.02$\pm$0.20\\
E8 & 12/12/2013 & 03:03 & W60S27 & B2.2 & 3.64 & 1.94$\pm$0.51\\
E9 & 01/08/2014 & 03:40 & W83N40 & M3.7 & 2.05 & 1.19$\pm$0.30\\
\hline
\end{tabular}
\caption{A list of studied event parameters. $Q_0$ and $Q_1$ denote the fitted intercept at 0.1~MeV/nuc and the power law indices of the observed quiet-time suprathermal Oxygen spectra, respectively (see text for details).}
\label{eventstable}
\end{table}

\subsection{Quiet-Time Suprathermal Particle Observations}
\begin{figure}[ht]
\includegraphics[width=1.0\columnwidth]{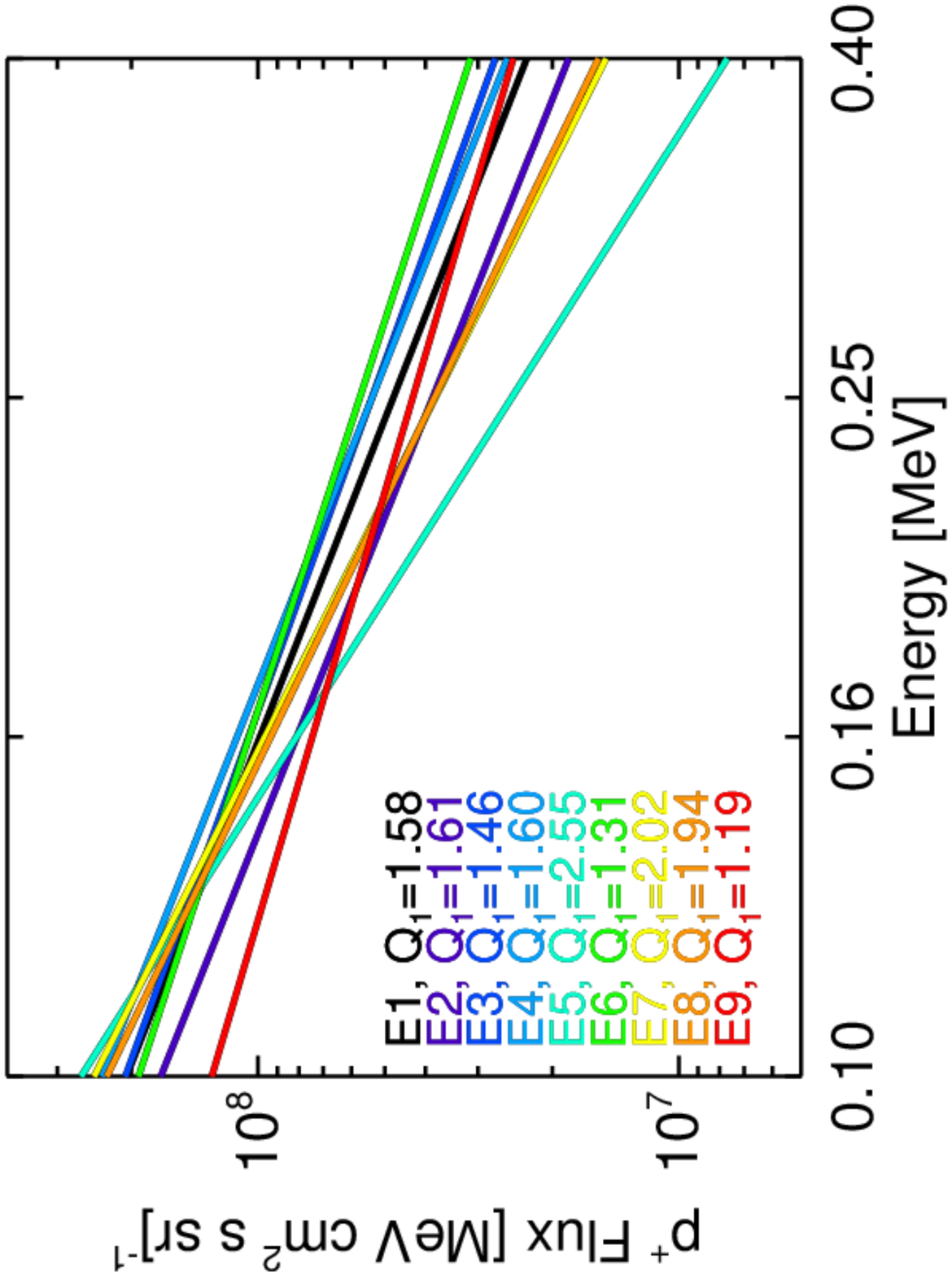}
\caption{Estimated low-coronal suprathermal proton seed spectra for all events, based on the in situ Oxygen observations near 1~AU. The colors correspond to individual events.}
\label{fig_input_spectra}
\end{figure}

In order to estimate the seed proton spectra, we used quiet-time Oxygen (O) in situ observations from ACE/ULEIS for periods immediately preceding (averaged over 24~hours) the arrival of the particles associated with the solar events. Figure \ref{fig_input_spectra} shows suprathermal observations for one of the events studied here. Panel a) shows Oxygen fluxes at ~0.05 - 8.7 MeV/n in 16 energy channels as measured by the ACE/ULEIS instrument \citep{Mason:1998}. Fig. \ref{fig_input_spectra} b) shows the magnetic field \citep[ACE/MAG]{Smith:1998} and solar wind speed \citep[ACE/SWEPAM]{McComas:1998} data for the same event. The sharp jump in the B and SW speed data indicates the shock arrival at the spacecraft (dashed vertical lines). Ambient and shock sampling time periods are also marked. Fig. \ref{fig_input_spectra} c) shows the energy spectrum of the event in the range E = 0.1-0.4~MeV/n, fitted by a power-law (blue curve) of the form $J = Q_0 E^{-Q_1}$, where $Q_0$ is the flux fit value at $E_0=0.1~MeV$, and $Q_1$ is the spectral index. The fuchsia curve indicates a power law fit to the event fluxes. 

Our choice of using the Oxygen spectrum is based on the assumption that spectral indices of most ions are similar (e.g., if coming from a diffusive acceleration process, a ubiquitous spectrum would result independent of the species). Furthermore, Oxygen presented a good tradeoff for the selected events between statistics, decent detection efficiency, and saturation issues during active periods for the low energy range covered in this study.


\subsection{Seed Spectra Estimation}

\begin{figure}[ht]
\includegraphics[width=1.0\columnwidth]{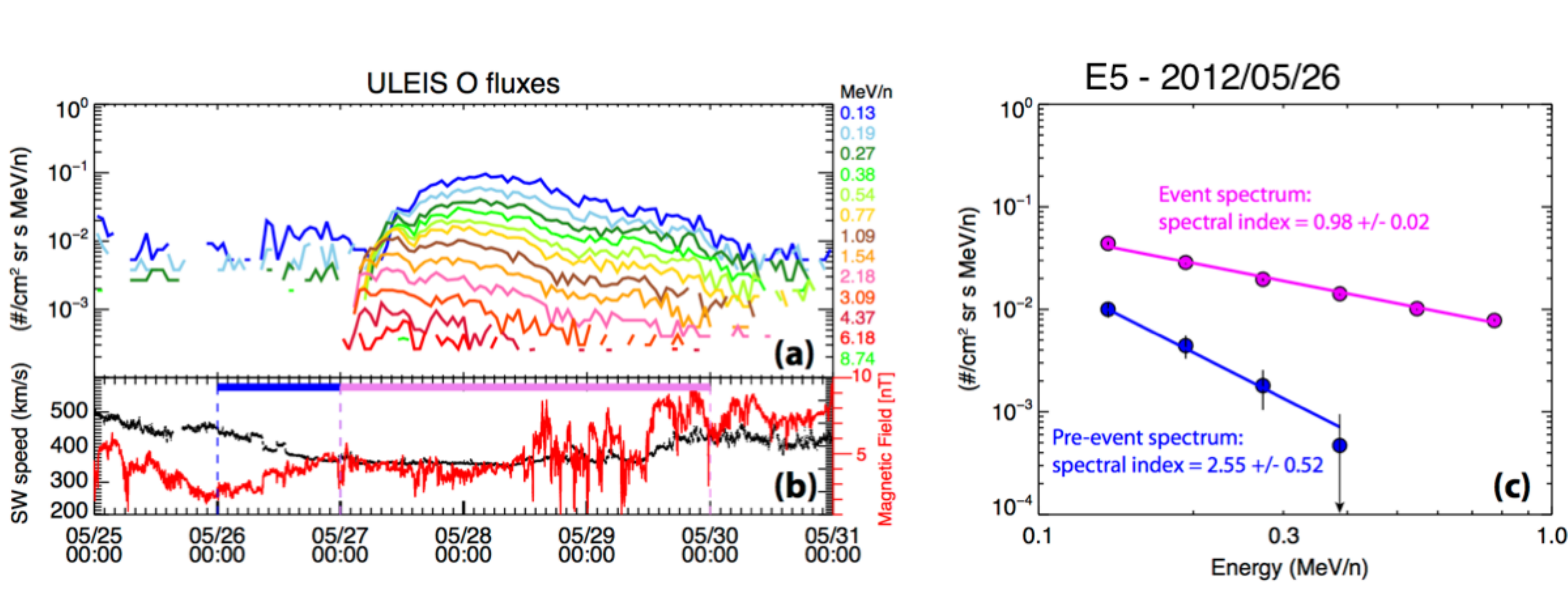}
\caption{Observed Oxygen fluxes from the ACE/ULEIS instrument for event E5. (2012/05/26). The plots show the energetic particle fluxes (panel \textbf{a)}), the solar wind speed and magnetic field strength (panel \textbf{b)}), as well as fitted spectra preceding and during the event (panel \textbf{c)}, blue and purple, respectively). The blue and pink horizontal lines in panel \textbf{b)} denote the periods of quiet-time and SEP event, respectively, used to form the average spectra in panel \textbf{c)}.}
\label{fig_observed_fluxes}
\end{figure}

The main assumptions in building the seed spectra near the Sun are: 1) The suprathermal fluxes observed at 1 AU are representative of quiet-time conditions in the low solar corona, and are formed by its ubiquitous distributed small-scale activity, rather than by large-scale eruptive events (flares or CMEs); 2) The suprathermal fluxes have not been heavily modified by magnetic effects during their transport between the Sun and 1 AU; 3) The Oxygen spectral slopes are similar to the proton spectra, for any given event; 4) Only the fluxes between 100~keV and 0.4~MeV are used; above 0.4~MeV, no source of energetic protons is assumed. With these assumptions, we use the power law fits to the observed averaged quiet-time spectra to reconstruct the suprathermal seed spectra, following the method of \citet{Kozarev:2013}. First, the O flux spectra observed at 1 AU are converted to proton spectra using abundances from \citet{Feldman:2003} and \citet{Reames:2014} (0.064$\pm$0.01\% relative O abundance). Then, they are scaled radially between 1 AU and 1.05~\rsun, assuming a simple inverse square relation with distance. The spectra are time-independent, representing the ambient state of the suprathermal proton distributions in the corona preceding the onset of each event. As the chosen events are all near the western limb, and are nominally well connected magnetically to Earth, we assume that these spectra are representative of the state of suprathermal charged particles in the corona, prior to the eruptions.

\section{Coronal Bright Front Analysis}
\label{s3}
\begin{figure}[ht]
\includegraphics[width=1.0\columnwidth]{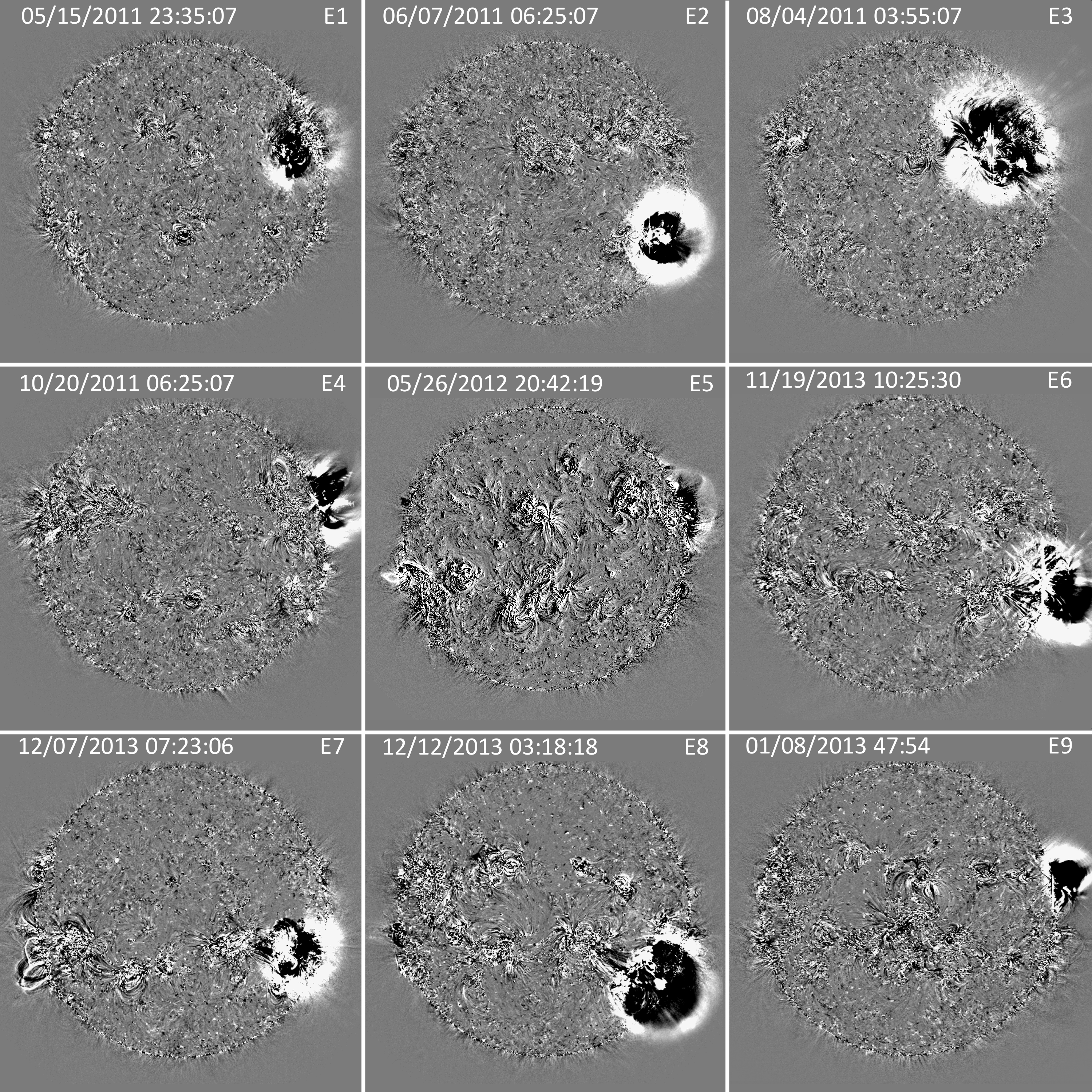}
\caption{Base difference SDO/AIA snapshot images, produced by subtracting a pre-event image of the corona from consecutive images, show the full-solar disk with already well-developed CBFs for all 9 analyzed events.}
\label{fig_euv_waves}
\end{figure}

The CASHeW framework's analysis tools were used to characterize the CBFs in all events, in order to derive the relevant parameters of the acceleration drivers (see details in \citet{Kozarev:2017}). Snapshots of the CBFs are shown in Fig. \ref{fig_euv_waves}. First, the kinematics of every CBF is estimated by automated fitting of time-dependent J-maps obtained in the radial direction from full-resolution, full-cadence 194~\AA-channel AIA images. Based on those, a coronal shock geometric surface (CSGS) model is constructed - a 3D spherical cap mesh surface with its center lying along the radial line passing through the eruption source on the solar disk. A limitation of this geometry is that it does not take into account the evolving shapes of CBFs. Real CBFs often exhibit an `over-expansion' in the lateral direction shortly after their formation, beyond a purely spherical expansion. Thus, CBFs often appear ellipsoidal, which may change the angles of interaction with the magnetic fields, and thus may affect the acceleration efficiency. It has been suggested that this over-expansion is due to the expansion of the developing CME flux rope/filament which drives the front \citep{Patsourakos:2010, Patsourakos:2012, Long:2017}. Here, we provide some information on whether lateral over-expansion relative to the radial evolution of each event is present, based on analyzing the lateral kinematics of all events at two heights in the low corona. This is summarized in Table \ref{latkintable}. It shows the average speeds in the poleward and anti-poleward lateral directions (parallel to the limb) from the radial direction of the event source at heights of roughly 1.08\rsun and 1.27\rsun. We find that some lateral expansion may have occurred in events E2, E3, E6, E7, and E8. The table also shows the maximum position of the front towards and away from the north pole at those two radial heights. For almost all events (with the exception of E3), there is good agreement on the final polar angles observed at the two radial heights. In future work, we will improve the geometrical description of the CBF model, relaxing the spherical assumption, in order to compare the effects of spherical vs. ellipsoidal geometry on the efficiency of acceleration.

A high-resolution Potential Field Source Surface (PFSS) model \citep{Schrijver:2003} of the coronal magnetic field is run using SDO/HMI data, for the time closest to, but preceding every event. The points of intersection between the CSGS model surface and the PFSS field lines at every timestep are determined, giving the upstream shock-field angle, $\theta_{Bn}$, and the magnetic field, $B$. The next step in the analysis is the estimation of density, density compression ratios, and temperature. Those are obtained by using the differential emission measure (DEM) model of \citet{Aschwanden:2013} and data from all six EUV channels of AIA, using the method of \citet{Vanninathan:2015}, as applied in \citet{Kozarev:2017}. For this work, we introduce a technique for obtaining DEMs quickly, which reduces the analysis time considerably. Rather than calculating DEM for all pixels in the image as in our previous work, we calculate it only for the pixels corresponding to the projections of the shock-field intersection points in the plane of the sky, as well as for the pixels immediately surrounding them. The average density value for pixels with minimized chi-squared DEM model fit is taken as the representative value. This method allows for much faster calculations of the model DEM. A full-FOV DEM model is calculated for all pixels of a pre-event image, to provide the pre-event density and temperature. The density jump is calculated as the ratio of the density at the cross-point pixels for a particular observation time to the pre-event density at the same locations.

\begin{figure}[ht]
\includegraphics[width=0.9\columnwidth]{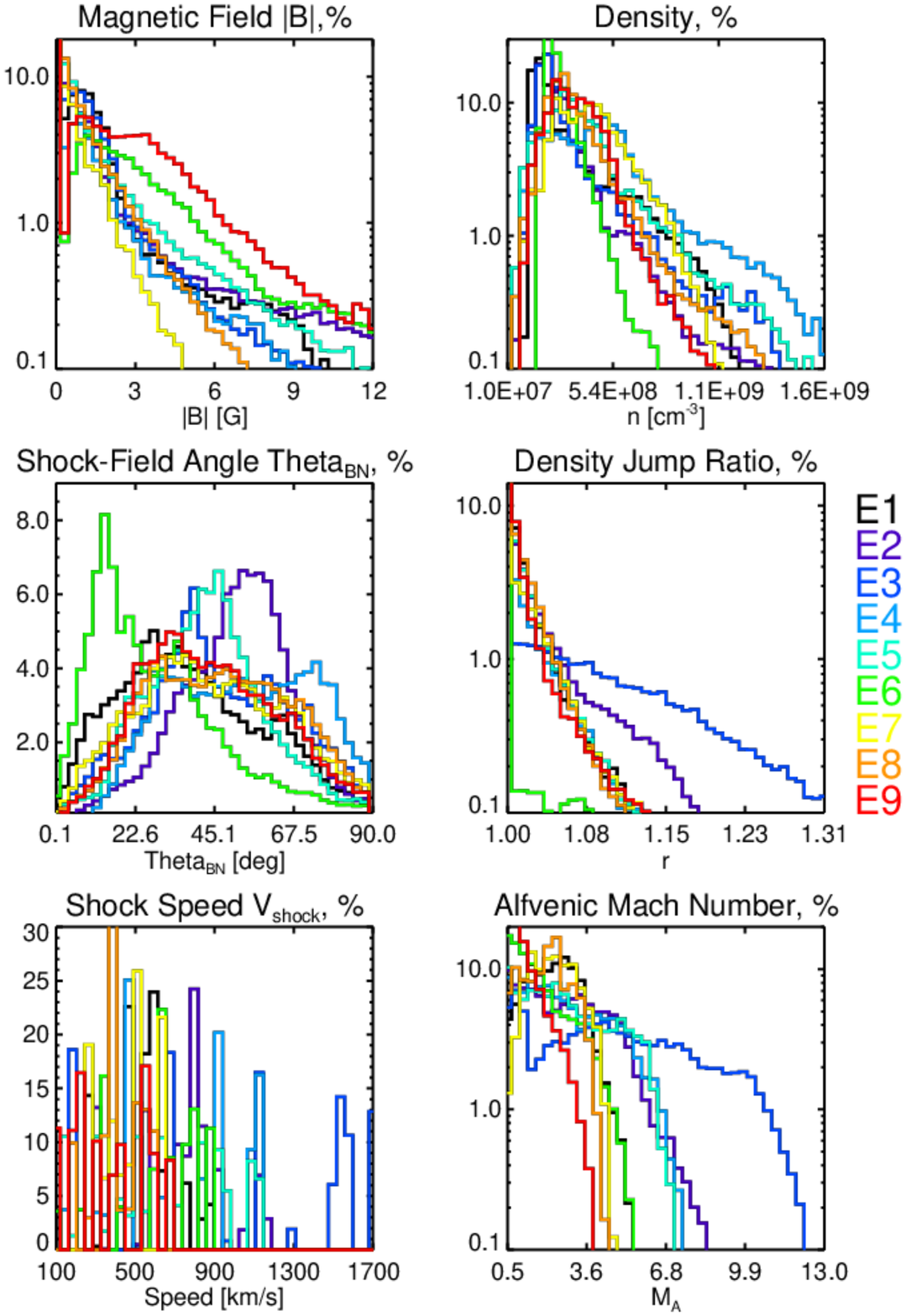}
\caption{Histograms of the main plasma parameters of the CBFs characterized with CASHeW. The colors correspond to the individual events, with the values drawn from for the overall evolution of each. The histograms are normalized to the sum of all values, and presented as percentages of the total.}
\label{fig_shock_parameters_histograms}
\end{figure}

Fig. \ref{fig_shock_parameters_histograms} summarizes the coronal conditions probed by the CBF in each event, as estimated with CASHeW. Histograms (with the number of bins set to 60 for all parameters) show the distributions of the upstream magnetic field magnitude, density, angle $\theta_{BN}$, density jump ratio, shock speed, and Alfv\'en Mach number, each normalized to the total for that distribution. The y-axes thus denote percentage fraction of the total distribution for each event. The distributions of magnetic field and density are similar in shape among the different events, with slight deviations at and below the 1\%\/bin level. On the other hand, for most parameters, the maximum of  The $\theta_{BN}$ distributions show some marked deviations for events E6 (peaking at very low values), and E2 and E4 (peaking at mid-and high values). The most significant deviations from the multi-event average are observed for the density jump ratio, for events E2 and E3, which exhibit the highest values of these parameters - again, below the 1\% level. The shock speed distributions are spiky due to the fact that for each time step of each event, all crossing points take a single speed (from the radial kinematics measurements). Finally, the Alfv\'en Mach number shows a wide variety of distribution shapes and maximum values, especially below the 1\% level. The highest values are for E3 and E2, and the lowest -- for E9. A relatively large portion of the E3 distribution is singled out at the high values beyond M$_A$=7. Based on the different distributions - especially those of $\theta_{BN}$, density jump, and M$_A$, we expect the strongest acceleration to occur for events E2, E3, and E4. The `disconnect' between shock speed and M$_A$ plots is due to the fact that the number of shock speed values is very limited (one speed is assigned to each front crossing at a particular time in each event) due to the way we model the CBF, while the density and magnetic field are different at each crossing and time, and thus have many more values represented in the histograms. This makes the M$_A$ histograms much smoother.

\begin{table}[ht]
\centering
\begin{tabular}{c c c c c c c c}
\hline\hline

Event \# & $R_{meas}$ & $\overline{V_{rad}}$ & $\overline{V_{lat,N}}$ & $\overline{V_{lat,S}}$ & PA$_{src}$ & max(PA$_{N}$) & max(PA$_{S}$)\\
 & \rsun & $km/s$ & $km/s$ & $km/s$ & \degrees & \degrees & \degrees \\
\hline
\multirow{2}{*}{E1} & 1.08 & \multirow{2}{*}{643.9$\pm1.5$} & 242.9$\pm17.4$ & 533.9$\pm27.8$ & \multirow{2}{*}{60.5} & 53.0 & 77.6 \\
 & 1.26 & & - & - &  & - & - \\
\hline
\multirow{2}{*}{E2} & 1.08 & \multirow{2}{*}{805.4 $\pm1.3$} & 791.6$\pm18.2$ & 628.2$\pm12.3$ & \multirow{2}{*}{120.0} & 80.7 & 152.1 \\
 & 1.28 & & 1327.1$\pm114.8$ & 1171.5$\pm305.3$ &  & 87.9 & 150.2 \\
\hline
\multirow{2}{*}{E3} & 1.08 & \multirow{2}{*}{1256.7$\pm4.4$} & 1633.6$\pm19.4$ & 863.9$\pm27.2$ & \multirow{2}{*}{64.6} & -0.9 & 102.7 \\
 & 1.28 & & 1731.6$\pm130.4$ & - &  & - & 136.6 \\
\hline
\multirow{2}{*}{E4} & 1.08 & \multirow{2}{*}{571.0$\pm1.0$} & 796.6$\pm113.1$ & 214.9$\pm18.4$ & \multirow{2}{*}{71.6} & 44.2 & 105.5 \\
 & 1.26 & & 737.8$\pm22.0$ & 599.5$\pm18.4$ &  & 42.7 & 105.5 \\
\hline
\multirow{2}{*}{E5} & 1.08 & \multirow{2}{*}{663.5$\pm1.8$} & 665.5$\pm9.4$ & 764.0$\pm9.1$ & \multirow{2}{*}{69.4} & 26.3 & 116.3 \\
 & 1.27 & & 883.6$\pm17.4$ & 2462.5$\pm32.1$ &  & 38.2 & 116.5 \\
\hline
\multirow{2}{*}{E6} & 1.08 & \multirow{2}{*}{498.0$\pm1.9$} & 649.4$\pm15.4$ & 533.0$\pm23.7$ & \multirow{2}{*}{110.7} & 82.7 & 142.7 \\
 & 1.26 & & 1075.9$\pm25.9$ & 825.1$\pm46.7$ &  & 86.7 & 131.3 \\
\hline
\multirow{2}{*}{E7} & 1.08 & \multirow{2}{*}{432.3$\pm1.2$} & 1603.1$\pm12.9$ & 1038.4$\pm17.1$ & \multirow{2}{*}{111.1} & 79.8 & 157.9 \\
 & - & & - & - &  & - & - \\
\hline
\multirow{2}{*}{E8} & 1.08 & \multirow{2}{*}{458.9$\pm0.9$} & 380.8$\pm5.1$ & 360.4$\pm2.1$ & \multirow{2}{*}{120.7} & 96.5 & 151.9 \\
 & 1.25 & &895.8$\pm3.7$ & 852.8$\pm8.0$ &  & 92.7 & 153.2 \\
\hline
\multirow{2}{*}{E9} & 1.08 & \multirow{2}{*}{442.4$\pm10.0$} & 496.1$\pm7.0$ & - & \multirow{2}{*}{76.3} & 58.8 & - \\
 & 1.25 & & 499.5$\pm24.4$ & 309.6$\pm53.1$ &  & 66.0 & 84.1 \\
\hline
\end{tabular}
\caption{Summary of lateral OCBF kinematics measurements, using front positions along lines of constant height above the solar limb towards and away from the solar north pole, starting at the source radial direction. $R_{meas}$ is the radial height of measurement in the sky plane; $\overline{V_{rad}}$ is the measured average radial speed, for reference; $\overline{V_{lat,N}}$ \& $\overline{V_{lat,S}}$ are the average plane-of-sky speeds in the lateral directions towards and away from the solar north pole, respectively; PA$_{src}$ is the polar angle of the radial direction of the source;  max(PA$_{N}$) \& max(PA$_{S}$) are the maximum front polar angle extents to the north and south from the radial direction, measured at the respective radial height.}
\label{latkintable}
\end{table}

\section{Particle Acceleration Modeling}
\label{s4}
We estimate the shock acceleration of the coronal protons using the diffusive shock acceleration (DSA) model introduced by \citet{Kozarev:2016}. It has been developed to easily incorporate the shock and field parameters derived by the CASHeW framework. The model solves analytically for the time-dependent proton distribution function at the shock front by ingesting the local shock speed, the local density and density change, the magnetic field magnitude upstream, as well as the shock-field angle, $\theta_{BN}$. A time-independent source distribution function is prescribed, as well as an upstream parallel scattering mean free path, which represents the transport conditions in the corona. Here, we have taken a constant value of 0.0055 \rsun, which is on the high end of the solar granule size distribution. The assumption is that the turbulence along magnetic field lines in the low corona is related to the macroscopic motions on the scales of photospheric convection \citep{Cranmer:2005}. The idea is that there is a characteristic scale of the magnetic turbulence on individual field lines, to which the charged particles react, and which is more or less constant within the limited radial extent of the model ($\approx$1.05-1.3\rsun). Of course, there is typically a spectrum of turbulent scales, which will be added in future versions of the model. This simplistic treatment will be relaxed in future studies.

In this work, we make a significant modification to the source particle distribution, described in previous work, by taking actual quiet-time suprathermal particle observations near 1 AU preceding each event as the basis of the input seed spectra in the model. The evolving information about the crossing points between the CSGS model and the PFSS field lines is directly input to the model for each timestep of each event. The acceleration of the input spectra is calculated for all field lines that cross the CSGS surface for at least three consecutive observational time steps. The particles are constrained to an individual field line for the duration of the process, and an escape condition is not implemented.

\begin{figure}[ht]
\includegraphics[width=1.0\columnwidth]{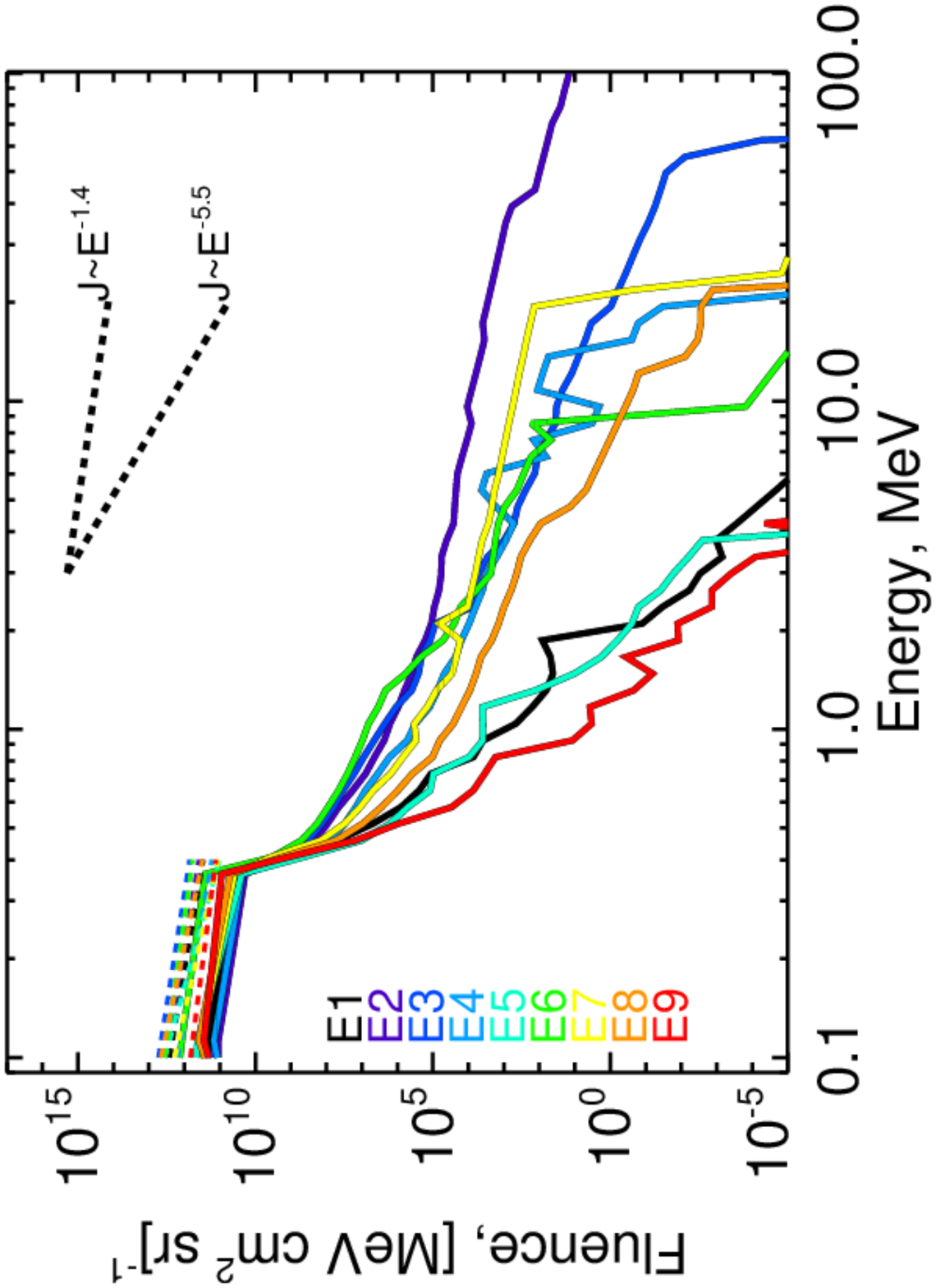}
\caption{Final SEP fluence spectra for each event, averaged over all field lines for which the acceleration was calculated. The input spectra are overlaid in dashed lines. The colors correspond to individual events.}
\label{fig_compare_dsa_fluences}
\end{figure}

Figure \ref{fig_compare_dsa_fluences} shows the proton fluences for all events, calculated from the final distribution function values, with different colors corresponding to different events. For each event, we constructed the average fluence by obtaining, for each final energy bin, the mean of the distribution of fluences for all shock-crossing field lines at that bin. We have controlled for outliers in these distributions by removing all points lying beyond 3$\sigma$ from the mean fluence value at each energy bin, and recalculating it with the remaining points. Note that the results shown in the plot are an average over both time and space for each of the compressive/shock waves under investigation. The colored dashed lines denote the source spectra for each event.

\begin{landscape}
\begin{figure}[ht]
\includegraphics[width=1.0\columnwidth]{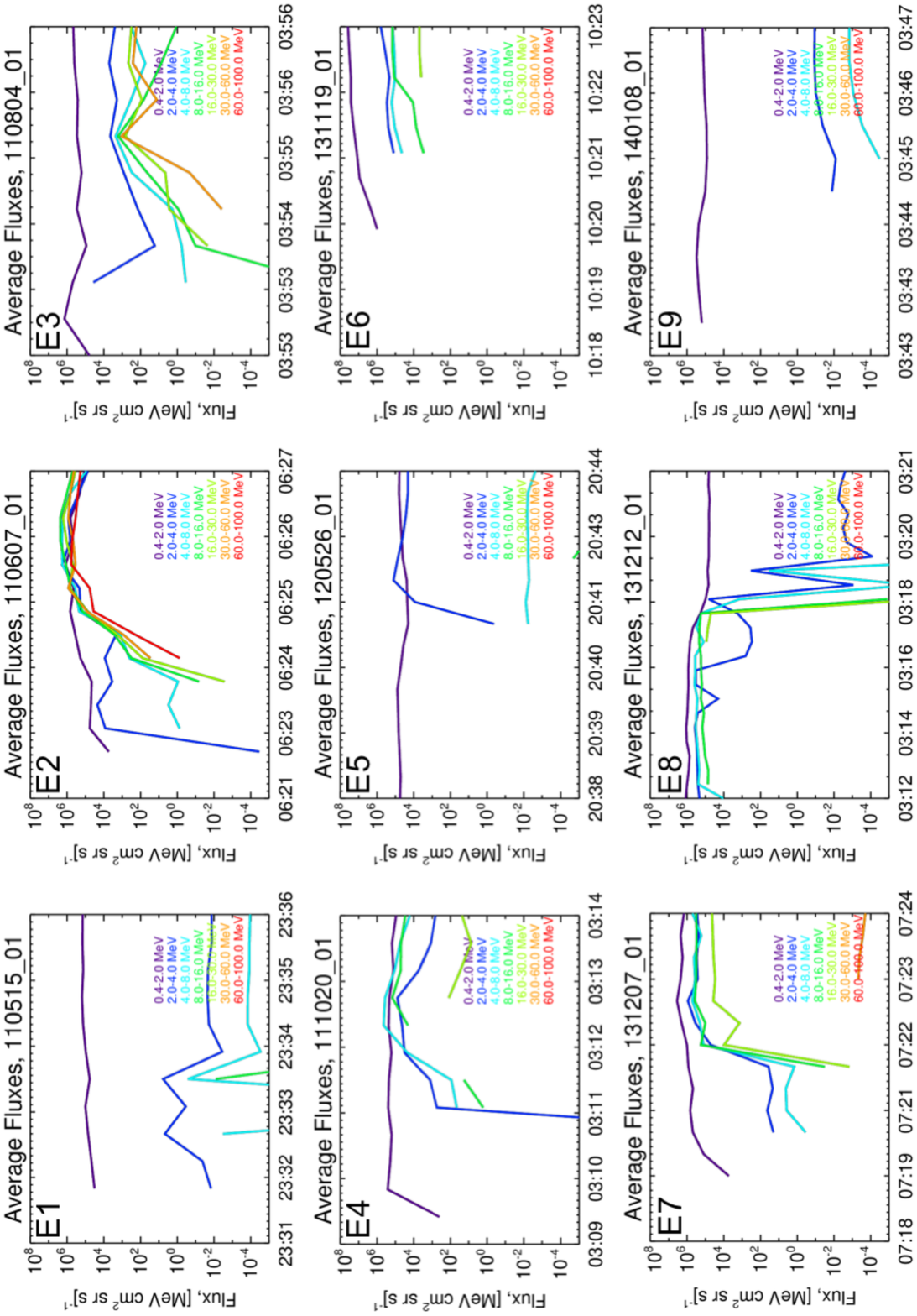}
\caption{A comparison of the average flux time series in 8 energy channels for all events, after outliers have been removed. The colors correspond to individual energy channels.}
\label{fig_compare_dsa_fluxes}
\end{figure}
\end{landscape}

The temporal evolution of the modeled proton fluxes separated into seven coarse energy `channels' for each event, averaged over all acceleration locations, is given in Figure \ref{fig_compare_dsa_fluxes}. The UT time is on the x-axis, while flux is on the y-axis. Fluxes at progressively larger energies are shown with lines colored from purple to red. Similarly to the fluences shown in Fig.  \ref{fig_compare_dsa_fluences}, the outliers in the flux distribution at each energy bin have been removed. As expected, the fluxes at the lower energies generally have higher values, and appear within the flux range before those at the higher energies. The top middle panel shows, for event E2, most clearly a time evolution of the fluxes, typical of SEP events. Similar evolution can also be observed for events E3, E4, and E7. For the other events, the acceleration is less and slower.

Table \ref{resultstable} summarizes the results from the DSA calculations. It includes the number of shock-crossing PFSS field lines, for which the DSA model is run, the maximum duration of acceleration, and the maximum energy reached for each event. The table also provides the final proton fluence spectral indices, as the shock front noses reach $\sim$1.3\rsun~projected height. The full spectral fit indices were calculated by fitting the final fluences shown in Fig. \ref{fig_compare_dsa_fluences} in the range 0.4-100~MeV. The column with header `0.1-2.0 MeV Spec. Index' represents the modeled power law spectral index in the range 0.1-2.0 MeV near the Sun while the shocks were within 1.3 solar radii, while the final column with header `1 au O Spec. Index' represents the spectral index measured in the range 0.1-2.0 MeV/n by the ULEIS instrument at 1 au, averaged over each event. We have provided the oxygen 0.1-2.0 MeV spectra both to provide a low energy range comparison for weak events, and for comparison between the model results and the ULEIS observations, as ULEIS pre-event suprathermal (0.1-0.4~MeV) spectra are used to derive the input spectra for driving the DSA model.

We note that at such an early stage of the eruptions, we do not expect a good match between the model results and observations, as the fluences depend strongly on the further dynamics of the source near the Sun, as well as on their transport to 1 AU; thus, we cannot validate the model results by matching them to observed spectra without proper modeling of the transport. Comparing the two sets of indices directly may give information of how much acceleration occurs in the earliest stages of the events, versus later on. This is the goal of future work.

\section{Discussion}
\label{s5}

A comparison between the overall distributions of the plasma parameters of the events (Fig. \ref{fig_shock_parameters_histograms}) and the final fluences (Fig. \ref{fig_compare_dsa_fluences}) shows a good correspondence between the highest energies reached in events E2 and E3, and the distributions of density jump ratio and M$_A$. For the rest of the events, a connection is not so clear. For example, the fluences of event E7 (yellow color) are relatively high and reach almost 30~MeV - however, the plasma parameters at the shock crossings for that event do not seem to warrant significant acceleration. Conversely, event E4 has high values in the \thetabn and M$_A$ distributions, but only weak acceleration, as judged by the fluences plot in Fig. \ref{fig_compare_dsa_fluences}.

\subsection{Influence of Seed Spectra}

\begin{figure}[ht]
\includegraphics[width=1.0\columnwidth]{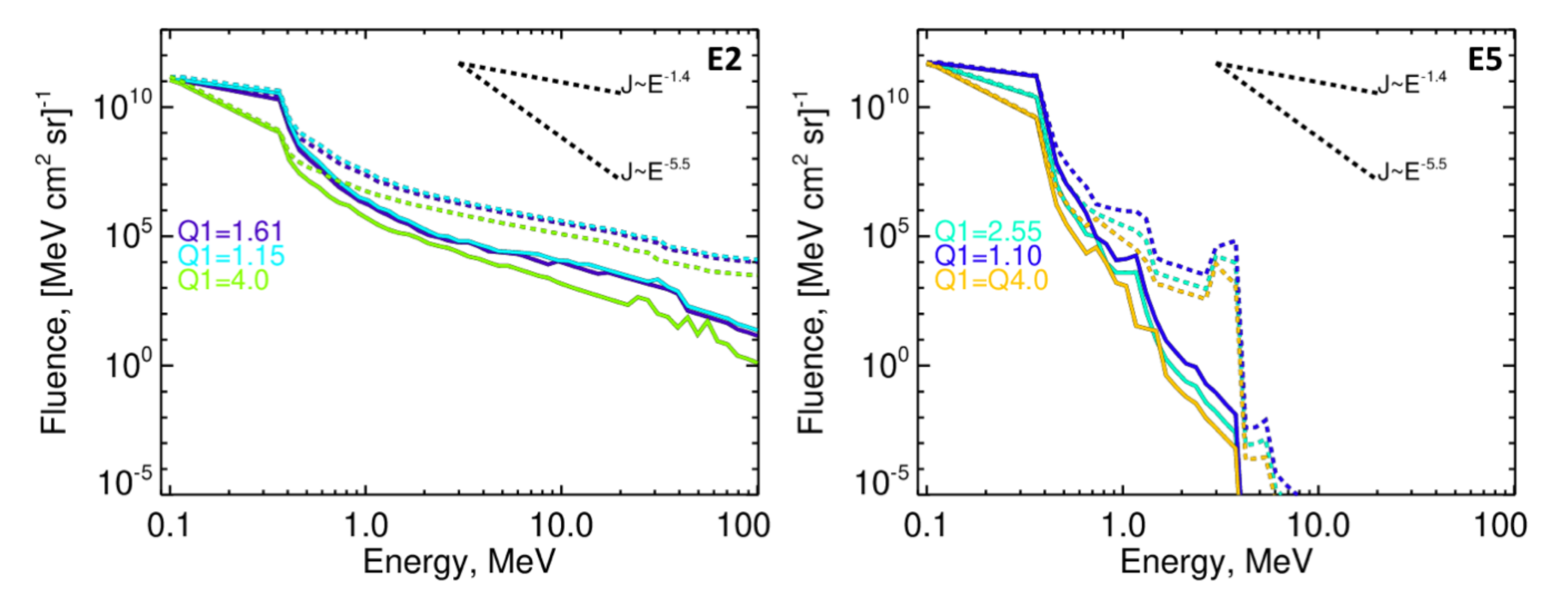}
\caption{(Left) Comparison of DSA model results for the strong event E2, using three different input spectral slopes (1.15, 1.61, 4.0), with the same set of shock plasma conditions. (Right) The same comparison for the weak event E5, with input spectral slopes 1.10, 2.55, and 4.0. The dashed lines are the fluences prior to removing outliers.}
\label{fig_fluences_multiseed_comparison}
\end{figure}

We have evaluated the influence of different input seed spectra for a given set of model results using three different input spectra for each of two of the events - one with strong (E2), and one with weak acceleration (E5). We have evaluated the changes to the final spectra by varying the input spectra slopes, while keeping the intercept the same. Figure \ref{fig_fluences_multiseed_comparison} shows the resulting fluences. In the left panel is the relatively strong event of June 07, 2011. It shows the final fluences from three different DSA model runs, with input slopes of 1.15, 1.61, and 4.0. The dashed lines are the fluxes before removing outliers. The difference between the steepest and flattest resulting spectra is about an order of magnitude throughout the spectra.

In the right panel of Fig. \ref{fig_fluences_multiseed_comparison} are shown the fluences for the relatively weak event on May 26, 2012, using the shock and plasma parameters for that event. The input slopes are 1.10, 2.55, and 4.0. The differences vary between less than a magnitude, and up to three magnitudes in the fluences over the energy range in question. Overall, the results show that there is an appreciable difference in the resulting fluences both for E2 and E5, in general to within one to three orders of magnitude. However, for the current list of events the spread in slope of the seed spectra is on the order of 0.5, so the variation should not cause a strong influence on the fluences.

\subsection{Influence of Inferred Plasma Parameters}
\begin{figure}[ht]
\includegraphics[width=1.0\columnwidth]{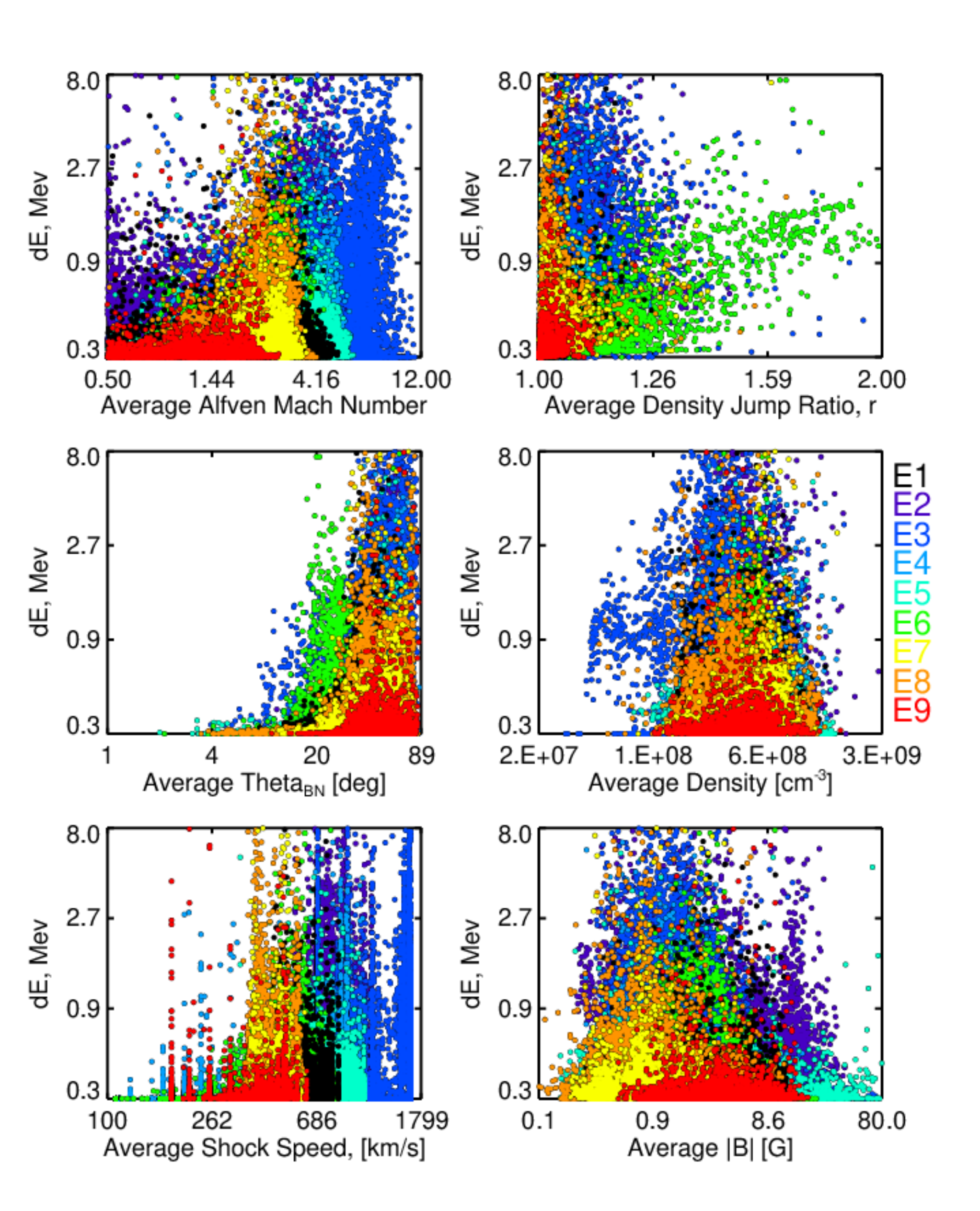}
\caption{Scatter plots of the change in energy achieved for all individual shock crossings of all events versus the average values of $M_A$, $r$, \thetabn, $n$, $V_{shock}$, $|B|$, over the acceleration period. Colors again correspond to individual events. The scaling on both axes for all panels is logarithmic.}
\label{fig_sep_shock_comparisons}
\end{figure}

We next look at how the average values of the plasma parameters at the shock crossings influence the energy gains of the protons. Fig. \ref{fig_sep_shock_comparisons} shows, for all events, scatter plots of the energy gained by the protons on each field line that crossed the shock and went through the acceleration process, versus the average shock crossing plasma parameters for that field line. Different colors correspond to different events. We have set the maximum energy change to a value that allows to explore the bulk behavior, leaving outliers out. In the case of average \thetabn (top left panel), most of the points lie beyond 15-20\degrees, and the distributions show a clear dependence on this parameter. As expected, the highest energy increases occur for the highest \thetabn values.

A similar behavior is seen for $V_{shock}$ (lower left panel) - more energization is seen where the average shock speeds are higher. We note that in this case there are discrete values of $V_{shock}$ due to the modeling setup, which can be seen in the plot. The higher average magnetic field magnitudes (upper right panel), on the contrary, seem to be loosely related to lower increases of the proton energies, although a very broad range of average $|B|$ values are represented. This may be due to early acceleration at locations of high $|B|$ and \thetabn, which is reduced as the compressive front enters regions of lower $|B|$ and more parallel \thetabn geometries. Finally, the average Alfv\'en Mach number (bottom right panel) shows that, for most events, distributions generally peak in energy near (but markedly below) the high values of M$_A$. A comparison with the average fluences for all events (Fig. \ref{fig_compare_dsa_fluences}) confirms the expected connection between the per-crossing energization and the overall fluences for events E2 and E3.


We find that the shocks do not produce significant energization of the protons for the vast majority of the shock-crossings, for the period under consideration. Of the 83056 field lines, for which we ran the DSA model, only 4533 (5.5\%) of the distributions at the crossings reach final energies beyond 1~MeV, and only 225 (0.27\%) - beyond 10~MeV. This is to be expected, given the short time for acceleration allowed, and the dynamic changes in the acceleration parameters of the compressive fronts. In general, we expect further acceleration higher in the corona.We are currently implementing an extension to the modeling chain, which will incorporate observations of the shock waves from LASCO, and will allow us to estimate SEP acceleration out to about 6\rsun. Results will be presented in a complementary follow-up paper.

\begin{table}[ht]
\centering
\begin{tabular}{c c c c c c c}
\hline
Event & \# Lines & Max. Acc. & Max. & Full & 0.1-2.0 MeV & 1 au O\\
 & & Duration & Energy & Spec. Index & Spec. Index & Spec. Index\\
\hline
E1 & 9434 & 4.0 min & 5.4 MeV & 21.5 & 10.6 & 1.8$\pm$0.04\\
E2 & 14801 & 4.8 min & 100.0 MeV & 2.6 & 5.5 & 1.31$\pm$0.04\\
E3 & 8980 & 2.4 min & 62.6 MeV & 8.3 & 6.1 & 1.59$\pm$0.04\\
E4 & 2796 & 3.2 min & 19.4 MeV & 5.4 & 6.8 & 1.45$\pm$0.04\\
E5 & 10229 & 6.0 min & 3.8 MeV & 17.3 & 10.9 & 1.01$\pm$0.04\\
E6 & 12780 & 5.2 min & 13.7 MeV & 8.4 & 6.8 & 2.04$\pm$0.05\\
E7 & 3898 & 4.0 min & 24.5 MeV & 6.1 & 6.9 & 3.11$\pm$0.05\\
E8 & 15573 & 9.2 min & 21.8 MeV & 9.0 & 8.0 & 1.73$\pm$0.04\\
E9 & 4565 & 4.0 min & 4.2 MeV & 20.5 & 13.6 & 1.36$\pm$0.02\\
\hline
\end{tabular}
\caption{A summary of the modeled DSA fluences and some observations, for all events. The columns represent the number of lines that the DSA model is run on, the longest acceleration process duration for that event, the maximum energy at fluences above 10$^{-5}$~[MeV~cm$^2$~sr]$^{-1}$, the overall spectrum (0.4-100~MeV) power law fit index, a low-energy range (0.1-2.0~MeV) model power law fit index, and the power law fit index to ULEIS 1~au observations over each of the events.}
\label{resultstable}
\end{table}


\section{Summary}
\label{s6}

In this work, we have presented a study of the coronal shock acceleration of protons in 9 western near-limb eruptive events. First, we estimated the relevant plasma parameters for the DSA process along the expanding low coronal shock surface fitted to EUV CBF observations, using the CASHeW framework. We find a variety of plasma parameters that vary significantly with time and position along the shock surface. Thus, there is no single dominant geometry/configuration of the eruptions, contrary to what has been suggested previously.

Next, we estimated quiet-time pre-event suprathermal Oxygen spectra from 1~AU observations with ACE spacecraft. We converted the Oxygen spectra to coronal proton spectra assuming a typical relative abundance of O to H, and inverse square radial dependence. The resulting source spectra were fed to the DSA analytic model of Kozarev \& Schwadron 2016, which calculated the overall time-dependent acceleration of protons for each of the 9 events. For the input parameters given, the model produced significantly accelerated resulting spectra. For six of the events, they reached energies above 10~MeV, and for two events - energies above 60~MeV. We found that for the mean free path used, most shocks can accelerate protons to over 10 MeV in 3-7 minutes. 

As expected, we found the strongest dependence of acceleration on angle \thetabn, shock speed, and Alfv\'en Mach number. We found a dependence on the input spectra, which affects the final fluence spectra for a given set of plasma/shock conditions. However, when comparing different events, we were not able to extract a correlation between the input spectra and the final fluences, due to the much varied coronal conditions from one event to another. In a follow-up study, we will extend the modeling to larger distances in the corona by incorporating lateral measurements of the CBF kinematics, as well as data-driven models for the plasma environment in the corona \citep{Zucca:2014}.\\


\acknowledgements
We thank an anonymous referee for providing useful comments and suggestions. Author A.F. acknowledges support provided by the Deanship of Scientific Research (DSR) at the King Fahd University of Petroleum and Minerals (KFUPM) for funding this work through the Project no. IN151005. Author M.A.D. acknowledges partial support by NSF SHINE grant AGS-1460118.

\bibliography{multievent_cashew_paper1}
\end{document}